\documentclass[twocolumn,twoside,slac_two]{revtex4}
\usepackage{graphicx}
\usepackage{bm}
\usepackage{fancyhdr}
\usepackage{amssymb}
\begin{document}

\title{Search for Sources of  Ultrahigh Energy Cosmic Rays}

\author{A.A. Mikhailov}
\thanks{Supported by grant Russian FBR 08-02-00437} 
\affiliation{Yu.G. Shafer Institute of Cosmophysical Research and Aeronomy, 31
Lenin Ave., 677980 Yakutsk, Russia} 



\begin{abstract}
The arrival directions of ultrahigh energy extensive air showers (EAS) by Yakutsk, AGASA and P. Auger data are considered. It is found that the arrival directions of EAS in the Yakutsk and AGASA data are correlated with pulsars from the side Input, and in the P. Auger data are correlated with pulsars from the  Output of the Local Arm of Orion. It is shown that the majority of these pulsars have a short  rotation period around their axes, than expected by the  pulsar catalogue. 

\end{abstract}

\maketitle
\thispagestyle{fancy}

\section{INTRODUCTION}

    First we  analyzed extensive air showers (EAS) from the data of the Yakutsk EAS array. Showers with energy $E>5\times10^{18}$ eV, with zenith angles  $<60^{\circ}$ and  axes lying inside the perimeter of the array are considered. The accuracy of defining the solid angles of arrival directions is $5-7^{\circ}$, and for energy - $\sim 30\%$. 

Earlier we have found showers with a deficit in the  muon content \cite{mikh1}. Theoretical calculations  \cite{chri} show that the  muon content of showers reflects the mass composition of the particles which formed them. Probably, EAS with the usual number of muons are formed by  charged particles, and EAS with a deficit by neutral particles.

\section{EXPERIMENTAL DATA}

Among EAS with a deficit in the number of muons we found 21 EAS without any muon component (the threshold for triggering the muon detectors   is $>1$ GeV). If the probability of registering these EAS was $>10^{-3}$ the given EAS was excluded from consideration. Each registered shower without a muon component was carefully checked - whether the measured parameters of the EAS are correct. 
We also  found 5 EAS with poor muon number - the muon density at distances $>100$ m from the axis was less than  expected by more than $3\sigma$. 

The celestial co-ordinate distribution of these 26  EAS is shown in Fig.1. The distribution of EAS with a deficit of muons is not isotropic, but some excess in the observed number of EAS is observed from the galactic plane: $n(|b|<30^{\circ})/n(|b|>30^{\circ}) = 1.9 \pm 0.7$. In the case of isotropy this ratio would equal 1.2 according to reference \cite{efim1}.

 Among these EAS we found 5 doublets of which 4 are located in one region of the  celestial sphere:  $\delta=20^{\circ}-75^{\circ}$ and $60^{\circ}<RA<80^{\circ}$. The fifth doublet has two EAS associated with it, one without muons and the other -   muon poor, and is located near the Input of the Local Arm of the  Orion Galaxy.

\begin{figure}[t]
\includegraphics[width=7cm]{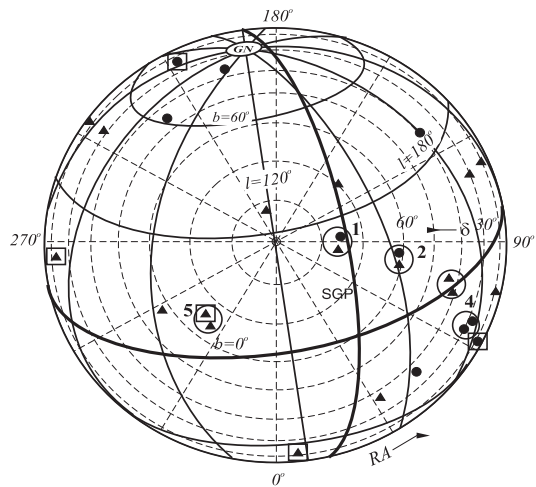}
\caption{The distribution of EAS with a deficit of muons: $\blacktriangle$ - EAS which are correlated with pulsars,  \textbullet - EAS which are uncorrelated with pulsars.  $\bigcirc$ - doublets,  $\Box$ - muon poor EAS. $\delta$ - declination, RA - a right ascension, b,l - galactic latitude, longitude.} \label{fig:1}
\end{figure}

We are interested in this maximum of doublets and consider the distribution of EAS with the usual number of muons as  a function of right ascension, RA. We divided the observed  energy region $E>10^{18}$ eV into 4 intervals: 
\newline  \indent 1) $10^{18} - 5\times10^{18}$ eV, 
\newline  \indent 2) $5\times10^{18} - 10^{19}$ eV, 
\newline  \indent 3) $10^{19} - 4\times10^{19}$ eV, 
\newline  \indent 4) $>4\times10^{19}$ eV 
\newline \noindent and the distribution of EAS in  right ascension was analyzed by the harmonic functions of Fourier (Fig.2). 

\begin{figure}[t]
\includegraphics[width=7cm]{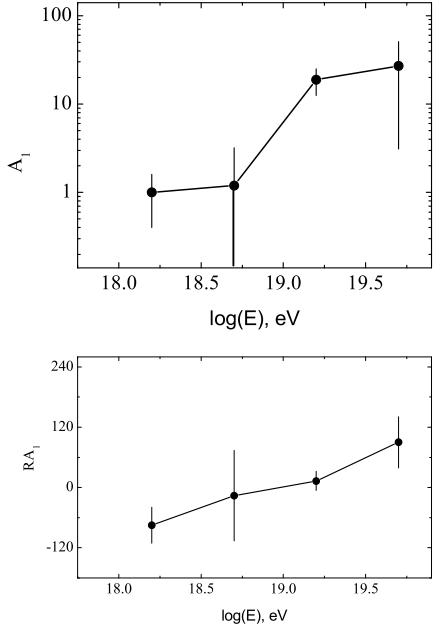}
\caption{Amplitudes $A_{1}$ and phase's $RA_{1}$ 1${^{st}}$ harmonic Fourier are 
shown in energy intervals.} \label{fig:2}
\end{figure}

\begin{figure}[htp]
\includegraphics[width=7cm]{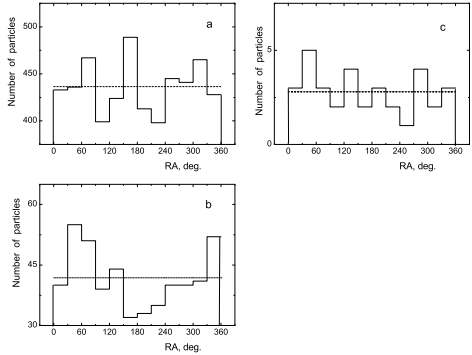}
\caption{Distribution of particles in energy intervals: a) $E=5\times10^{18} - 10^{19}$ eV;  b)$10^{19} - 4\times10^{19}$ eV; c) $E>4\times10^{19}$ eV.} \label{fig:3}
\end{figure}

\begin{figure}[htp]
\includegraphics[width=7cm]{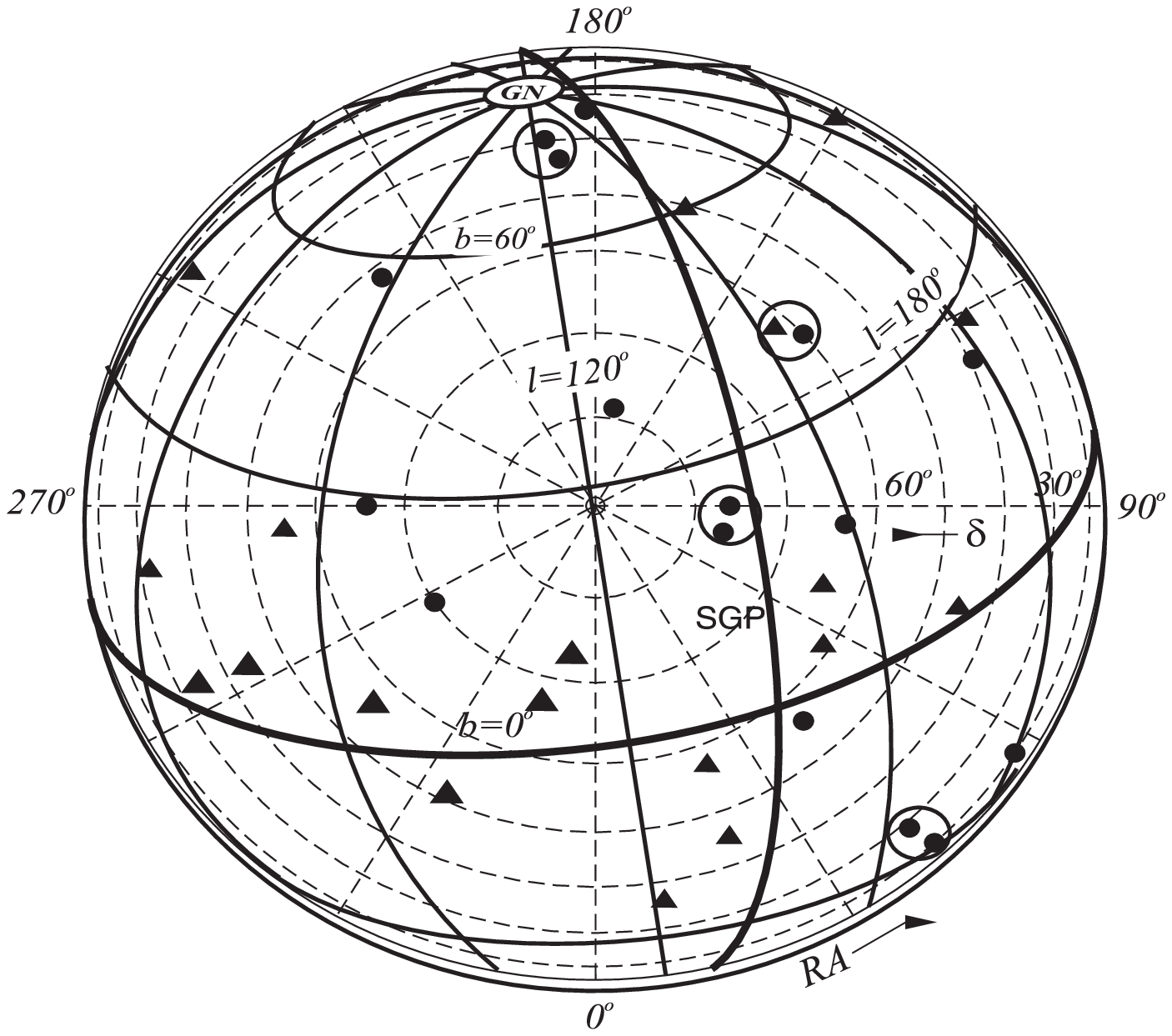}
\caption{Yakutsk: the distribution of particles with $E>4\times10^{19}$ eV: $\blacktriangle$, \textbullet - EAS which are correlated, uncorrelated with pulsars according to the pulsar catalogue \cite{tayl}.} \label{fig:4}
\end{figure}

\begin{figure}[htp]
\includegraphics[width=7cm]{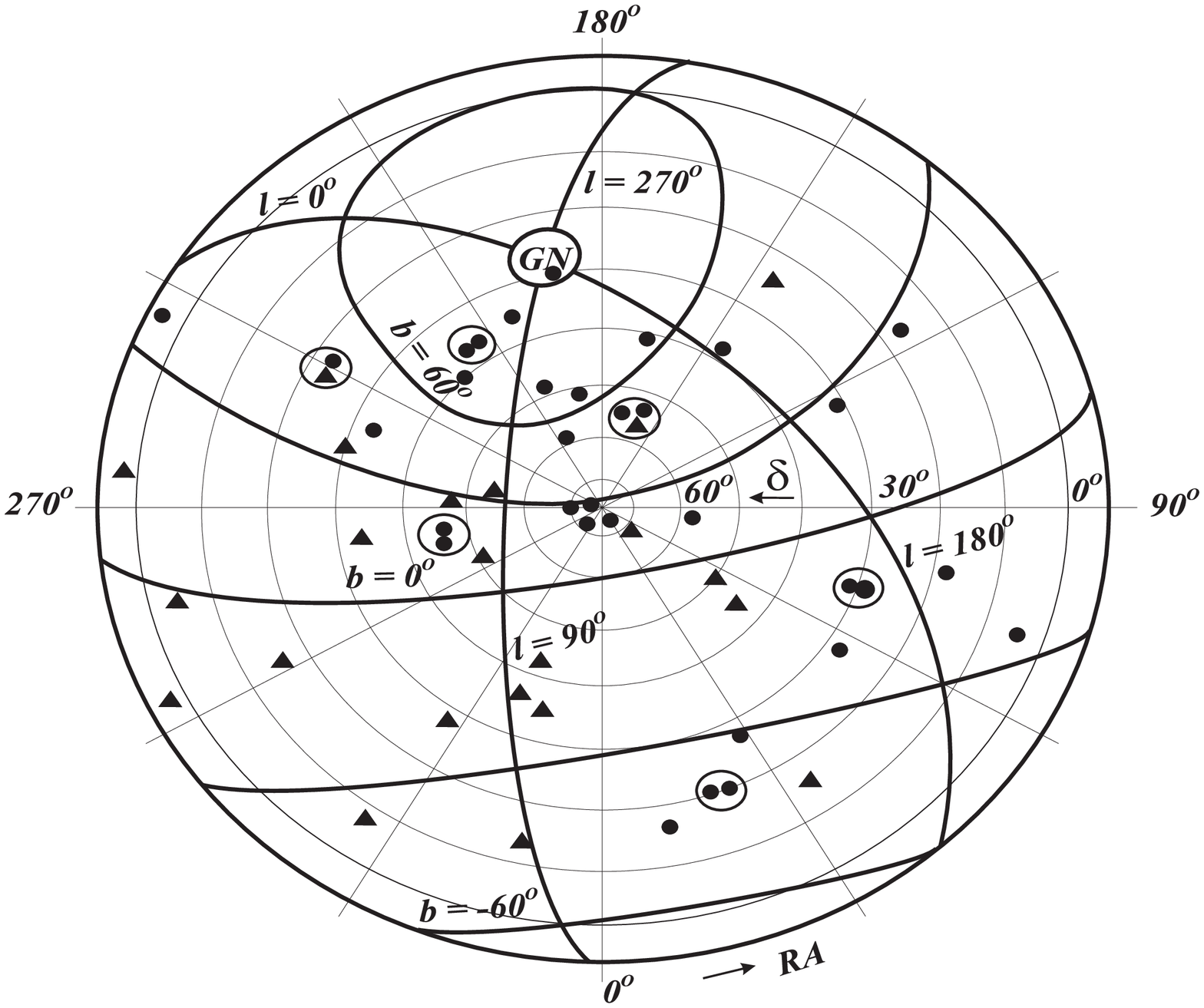}
\caption{AGASA: distribution of particles with $E>4\times10^{19}$ eV. $\blacktriangle$, \textbullet - EAS which are correlated, uncorrelated with pulsars according to the pulsar catalogue \cite{tayl}. } \label{fig:5}
\end{figure}

\begin{figure}[htp]
\includegraphics[width=7cm]{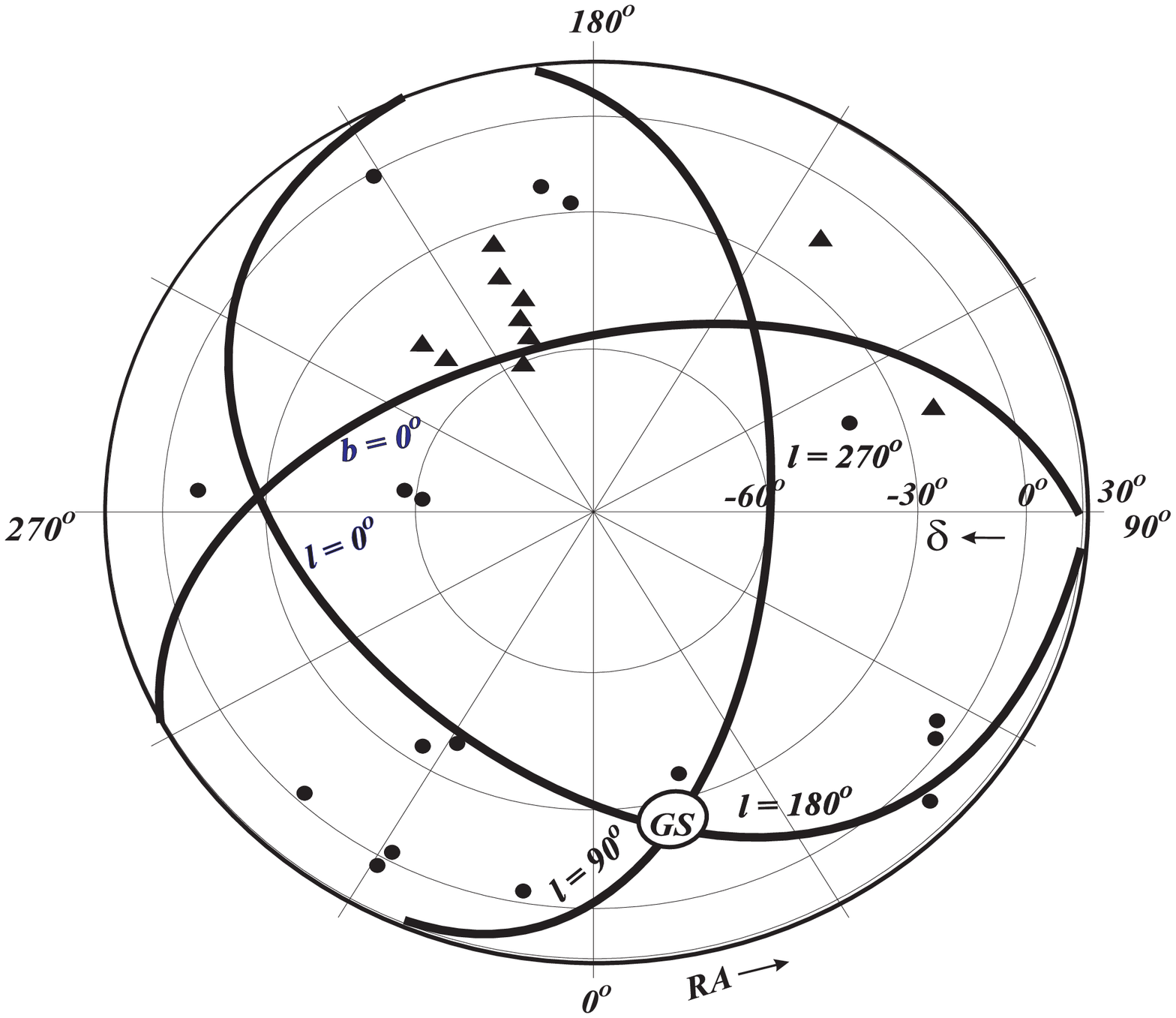}
\caption{P. Auger: distribution of particles with $E>4\times10^{19}$ eV. $\blacktriangle$, \textbullet - EAS which are correlated, uncorrelated with pulsars according to the pulsar catalogue \cite{tayl}.  } \label{fig:6}
\end{figure}

\begin{figure}[htp]
\includegraphics[width=7cm]{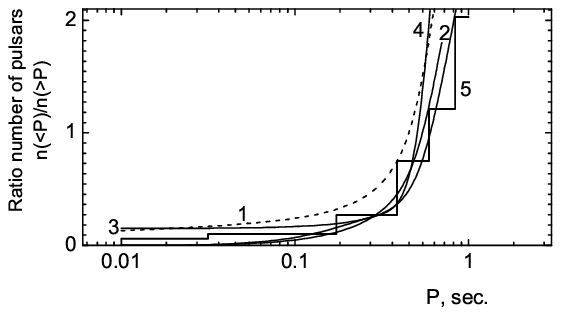}
\caption{ The ratio of the number of pulsars with period $P_{0}$  -  $n(P_{0}<P)/(n(P_{0}>P)$: 1 - pulsars, which are correlated with  muon poor EAS of Yakutsk; 
                             2 - pulsars,  correlated with EAS of Yakutsk; 
                             3 - pulsars,  correlated with EAS of AGASA; 
                             4 - pulsars,  correlated with EAS of P. Auger; 
                             5 - pulsars according to the catalogue \cite{tayl}.} \label{fig:7}
\end{figure}

Note, that the phase of the 1${^{st}}$ harmonic $RA\sim 300^{\circ}$ at $E\sim 10^{18}$ eV from the Local Arm of the Galaxy varies gradually with energy to
  $RA\sim 90^{\circ}$ at $ E \sim 4\times 10^{19}$ eV where 4 doublets are located. 

Further we also considered the distribution of EAS in right ascension. We divided the energy region 
into 3 intervals: 
\newline \indent 1) $5\times10^{18} - 10^{19}$ eV, 
\newline \indent 2) $10^{19} - 4\times10^{19}$ eV, 
\newline \indent 3) $>4\times10^{19}$ eV.

The distribution of particles is shown in Fig.3. We observe a maxima of the particle distribution  for the 2 first energy intervals at  $60^{\circ}<RA<90^{\circ}$. Most likely we observe neutral and charged particles from this region of the celestial sphere. 

From the 26 EAS with a deficit of muons only 17 EAS correlate with pulsars within an angular distance of $6^{\circ}$ \cite{tayl} (we choose $6^{\circ}$  because at  $E\sim10^{19}$ eV and at this angular distance from pulsars the  correlation between the arrival direction of EAS with the usual number of muons and pulsars  was maximum \cite{mikh2}). The arrival directions of these 17 EAS are marked by triangles. The distribution of these EAS which correlate with pulsars is mainly isotropic, but the majority of them are observed near the galactic plane. 

Further we  considered the arrival directions of EAS with energy $E>4\times10^{19}$ eV - are they correlated with pulsars? Thus, we  selected pulsars which are situated at angular distances $<6^{\circ}$ from the arrival directions of EAS. From the Yakutsk data we  found  19 such EAS from 34 (these EAS are noted by triangles in Fig.4), from the AGASA array \cite{haya} - 21 EAS from 57 (Fig.5), and from the P. Auger array \cite{auger} - 10 EAS from 27 (Fig.6). The arrival directions of these EAS, which are correlated with pulsars, are situated near the galactic plane and at Input (Yakutsk and AGASA) and at Output (P. Auger) of the Local Arm of the  Orion Galaxy.  Note, earlier we found an anisotropy in the arrival directions of particles with energy $E>4\times10^{19}$ eV from the side Input and Output of the Local Arm from the data of these arrays \cite{mikh3}. It is not possible to explain the correlation with pulsars and the anisotropy arrival directions of EAS  by an extragalactic origin of particles.

We considered the rotation periods of pulsars which are correlated with EAS. The ratio of the number of pulsars with periods $P_{0} <P$ to the number of pulsars which have  periods $P_{0}>P$ is shown in Fig.7 (for the Yakutsk array we  considered EAS with a deficit and with the usual number of muons). As seen in  Fig.7 the majority of pulsars have shorter periods, $P_{0}$, than   expected according to the  pulsar catalogue. Some authors have shown that short period pulsars can accelerate heavy nuclei up to $10^{20}$ eV \cite{blas,gill}.

\vspace{5mm}
\section{CONCLUSION}

The analysis of EAS from the Yakutsk data having a deficit in muon content   shows that a third  of them form doublets which are located  mainly at right ascension $60^{\circ}<RA<90^{\circ}$. At these coordinates a maximum distribution of normal EAS at energy $E\sim10^{19}$ eV is observed in the  Yakutsk data. 

 It is found that particles with energy $E>4\times10^{19}$ eV in the Yakutsk data, and  AGASA and P. Auger data are correlated with pulsars which are situated near the Input and Output of the Local Arm of the Orion Galaxy. The majority of these pulsars have a short  rotation period around their axes. The anisotropy and correlation of EAS with pulsars from the side Local Arm  of the Orion Galaxy  is difficult to explain by an extragalactic origin of cosmic rays. Most likely cosmic rays are of galactic origin and their sources are pulsars.

This paper has been supported by RFBR (project N 08-02-0497).

\end{document}